\documentclass[aps,prl,twocolumn,showpacs,groupedaddress]{revtex4-1}  
\usepackage{graphicx, amsmath}  

\begin{document}


\title{Observation of an Alfv{\'e}n Wave Parametric Instability in a Laboratory Plasma}


\author{S.~Dorfman}
\author{T.~A.~Carter}
\affiliation{University of California Los Angeles, Los Angeles, California 90095, USA}
\vskip 0.25cm


\date{\today}

\begin{abstract}
A shear Alfv{\'e}n wave parametric instability is observed for the first time in the laboratory.  When a single finite $\omega/\Omega_i$ kinetic Alfv{\'e}n wave (KAW) is launched in the Large Plasma Device above a threshold amplitude, three daughter modes are produced.  These daughter modes have frequencies and parallel wave numbers that are consistent with copropagating KAW sidebands and a low frequency nonresonant mode.  The observed process is parametric in nature, with the frequency of the daughter modes varying as a function of pump wave amplitude.  The daughter modes are spatially localized on a gradient of the pump wave magnetic field amplitude in the plane perpendicular to the background field, suggesting that perpendicular nonlinear forces (and therefore $k_{\perp}$ of the pump wave) play an important role in the instability process.  Despite this, modulational instability theory with $k_{\perp}=0$ has several features in common with the observed nonresonant mode and Alfv{\'e}n wave sidebands.
\end{abstract}


\maketitle


Alfv{\'e}n waves, a fundamental mode of magnetized plasmas, are
ubiquitous in space, astrophysical, and laboratory plasmas.  While the linear behavior of these waves
has been extensively studied~\citep{alfven42,wilcox60,morales97,gekelman99,vincena04}, nonlinear effects
are important in many real systems, including the solar wind and solar
corona.  Theoretical predictions show that these Alfv{\'e}n waves may be unstable to various parametric instabilities (e.g., Refs.~\citep{wong86, hollweg94, voitenko98}) even at very low amplitudes ($\delta{B}/B<10^{-3}$).  Parametric instabilities could contribute to coronal heating~\citep{pruneti97}, the observed spectrum and cross-helicity of solar wind turbulence~\citep{inhester90,zanna01,yoon08}, and damping of fast magnetosonic waves in fusion plasmas~\cite{lee98, oosako09}.

An abundance of theoretical work~\citep{sagdeev69,hasegawa76,goldstein78,wong86,longtin86,hollweg93,hollweg94} has found three types of parametric instabilities for a $k_\perp=0$ Alfv{\'e}n wave: decay, modulational, and beat.  The decay instability is the most widely known and involves the decay of a forward propagating Alfv{\'e}n wave into a backward propagating Alfv{\'e}n wave and a forward propagating sound wave.  By contrast, the modulational instability results in forward propagating upper and lower Alfv{\'e}nic sidebands as well as well as a nonresonant acoustic mode at the sideband separation frequency.  To allow the forward propagating waves to interact, the pump wave must be dispersive-- therefore the modulational instability at $k_\perp=0$ requires finite $\omega/\Omega_i$ through inclusion of Hall effects~\citep{hollweg94}.  Ponderomotive coupling between the pump and sideband Alfv{\'e}n modes self-consistently drives the nonresonant density perturbation parallel to the background magnetic field.  In this context, ``nonresonant'' means that the mode does not satisfy a dispersion relation in the absence of the instability drive; this is also called a quasimode in the fusion community~\citep{porkolab78, takase85}.

Both shear Alfv{\'e}n wave decay and modulational instabilities have been produced in numerical simulations~\citep{ghosh93, zanna01, matteini10a, verscharen12, gao13}, but observational evidence is limited.  Observations in the ion foreshock region upstream of the bow shock in the Earth's magnetosphere have found cases where a decay instability is possible, but results are not conclusive due to limited available data~\citep{spangler97, narita07}.

In this Letter, the first laboratory observations of a shear Alfv{\'e}n wave parametric instability are presented.  A single finite $\omega/\Omega_i$, finite $k_\perp$ Alfv{\'e}n wave is launched, and three daughter waves are observed when the amplitude of the pump is above a threshold: two sideband Alfv{\'e}n waves copropagating with the pump and a low frequency nonresonant mode.  Frequency and parallel wave number matching relations are satisfied.  Although these features of the observed instability are consistent with the $k_\perp=0$ modulational instability theory, the theoretical growth rate is too small to explain observations.  The spatial pattern of the daughter modes suggests a perpendicular (to the background magnetic field) nonlinear drive.

\begin{figure}[tbp]
\centering
\includegraphics[width=\columnwidth]{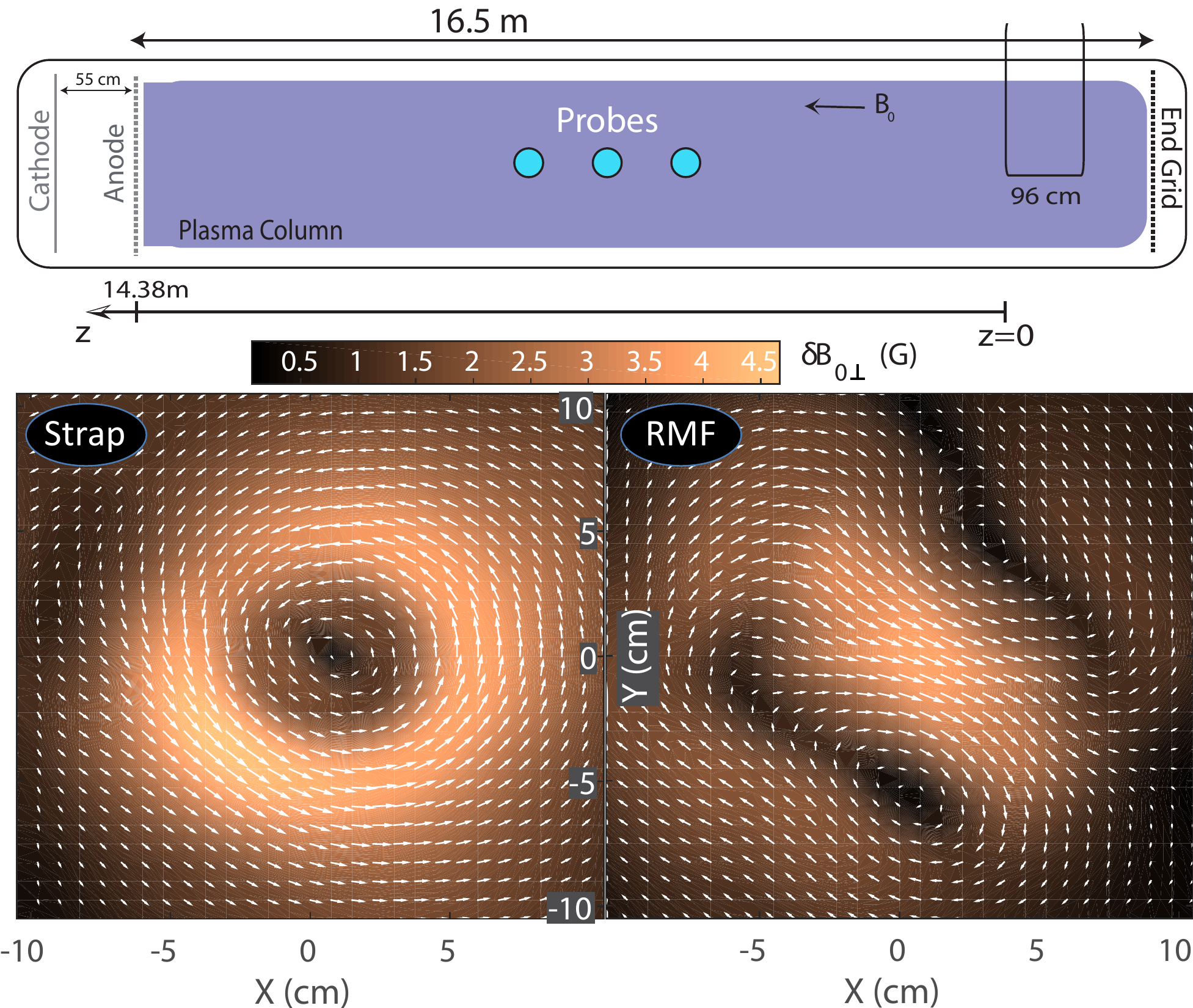}
\caption{Experimental setup in LAPD.  Top: An Alfv{\'e}n wave antenna on the right end of the device launches the pump wave.  Magnetic and Langmuir probes used to diagnose the interaction are shown.  Bottom: Spatial pattern of the pump wave in the $xy$ plane measured by a magnetic probe at $z=2.6$~m for the strap antenna (left, $B_0=1135$~G) and RMF antenna (right, $B_0=993$~G).}\label{fig:esetup}
\end{figure}

Experiments are conducted using the Large Plasma Device (LAPD) at UCLA, a cylindrical
vessel capable of producing a $16.5$~m long, quiescent, magnetized plasma
column for wave studies.  The BaO
cathode discharge lasts for $\sim{10}$~ms, including a several millisecond-long
current flattop.  Typical plasma parameters for the present study are
$n_e\sim10^{12}$~cm$^{-3}$, $T_e\sim5$~eV, and $B_0\sim1000$~G
($\beta \sim 10^{-3}$--$10^{-4}$) with a fill gas of helium.  Extensive
prior work has focused on the properties of linear Alfv{\'e}n waves
\citep{leneman99,gekelman00,vincena04,palmer05}.  Studies of the
nonlinear properties of Alfv\'{e}n waves have also been performed on the
LAPD; in these experiments, two launched Alfv{\'e}n waves nonlinearly interact to drive a nonresonant mode~\citep{carter06}, a drift wave~\citep{auerbach10}, an acoustic mode~\citep{dorfman13a,dorfman15}, or an Alfv{\'e}n wave~\citep{howes12}.

\begin{figure}[tbp]
\centering
\includegraphics[width=\columnwidth]{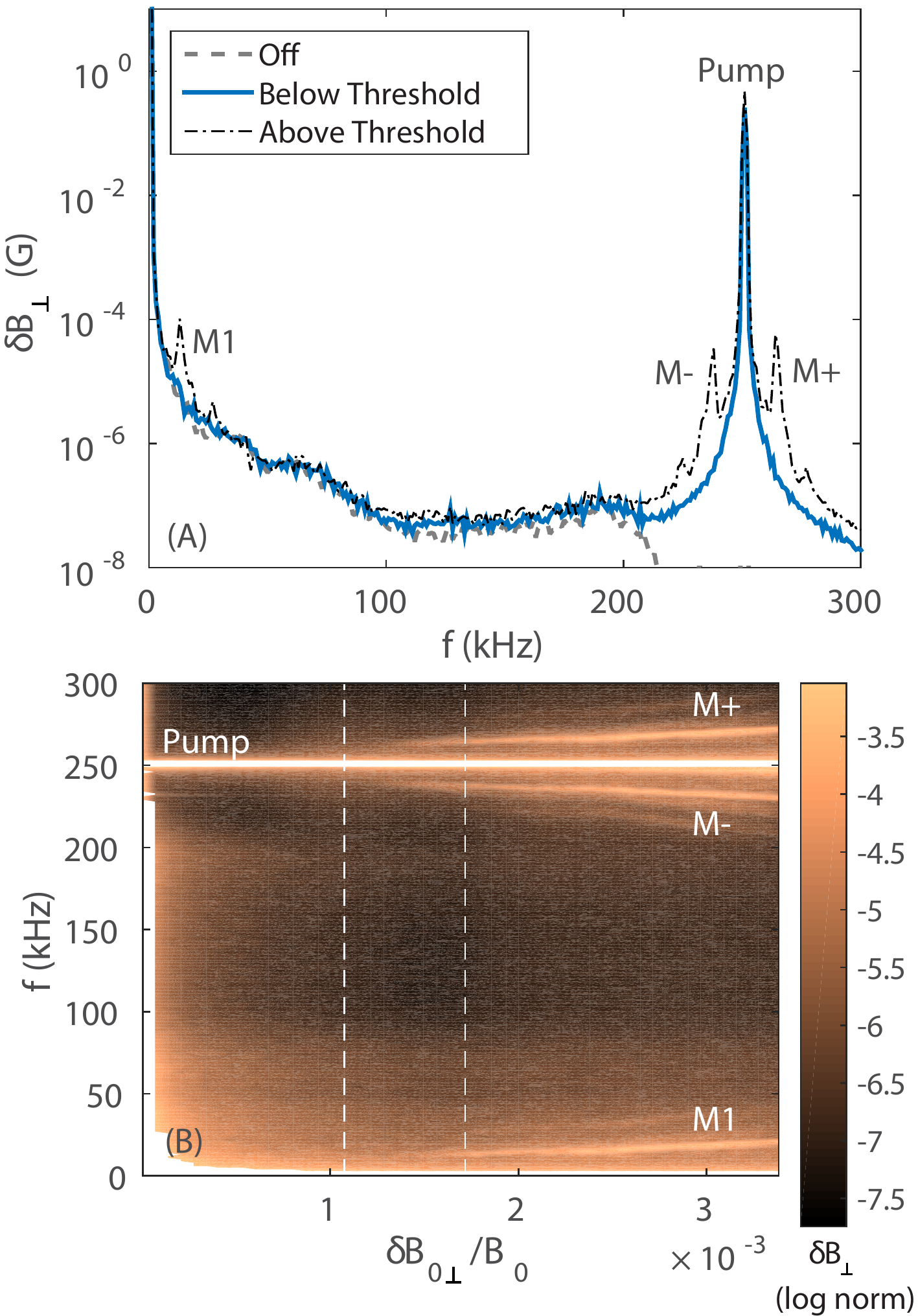}
\caption{Observed kinetic Alfv{\'e}n wave (KAW) parametric instability showing threshold behavior and parametric dependence.  RMF antenna, RHCP mode, $B_0=993$~G.  (a) Frequency spectrum from a magnetic probe at $x=0$, $y=-6$~cm, $z=2.6$~m for three pump mode amplitudes.  When the pump amplitude is above threshold for instability, three daughter modes are seen.  (b) Parametric dependence of the daughter mode frequency as a function of pump amplitude $\delta{B_{0\perp}}/B_0$.  The pump amplitude is $0$ on the log$_{10}$ color scale.  White vertical dashed lines represent values of pump amplitude from (a).}\label{fig:thpm}
\end{figure}

For the present set of experiments, a single antenna is placed at the far end of the LAPD, as shown in the top panel of Fig.~\ref{fig:esetup}.  This is either the $96$~cm long strap antenna~\citep{vincena13} shown in the diagram or the rotating magnetic field (RMF) antenna described in \citet{gigliotti09}.  The pump wave is launched at $\omega_0 \sim 0.67 \Omega_i$, producing the pattern in the plane perpendicular to $B_0$ shown for each antenna in the bottom panel.  The strap antenna launches a linearly polarized $m=0$ Alfv{\'e}n wave cone ($k_{\perp0}\rho_s=0.11$) in which oscillating magnetic field vectors (white arrows) circle the field-aligned wave current.  By contrast, the RMF antenna is set up to produce two field-aligned current channels ($k_{\perp0}\rho_s=0.21$) rotating around $B_0$ in an $m=1$ pattern \citep{gigliotti09}.  The rotation direction and hence wave polarization may be controlled by varying the antenna phasing.  To ensure the launched wave remains nearly monochromatic, the antenna current is digitized (not shown) and found to contain no significant sideband component.

In the plasma column in front of the antenna, magnetic and Langmuir probes detect the signatures of the pump and daughter modes.  Each probe is mounted on an automated positioning system that may be used to construct a 2D profile in the $x$-$y$ plane averaged across multiple discharges.

When the pump wave amplitude exceeds a threshold value, additional peaks are observed in the frequency spectrum, as shown in Fig.~\ref{fig:thpm}.  Panel~(a) of the figure shows the appearance of three modes: a low frequency mode ($M1$), a lower sideband mode ($M-$), and an upper sideband mode ($M+$).  The frequency matching relations $\omega_{\pm}\mp\omega_1=\omega_0$ hold.  However, $M1$ is not purely a density perturbation as predicted by the $k_\perp=0$ modulational instability theory; as seen in Fig.~\ref{fig:thpm}, the mode has significant magnetic character.

A clear parametric dependence of the mode frequencies on pump amplitude is shown in panel~(b) of Fig.~\ref{fig:thpm}.  As the pump amplitude $\delta{B_{0\perp}}/B_0$ increases above threshold, the frequencies of $M1$ and $M+$ increase; there is a corresponding decrease in the frequency of $M-$ such that frequency matching relations are satisfied at all wave powers.

\begin{figure}[tbp]
\centering
\includegraphics[width=\columnwidth]{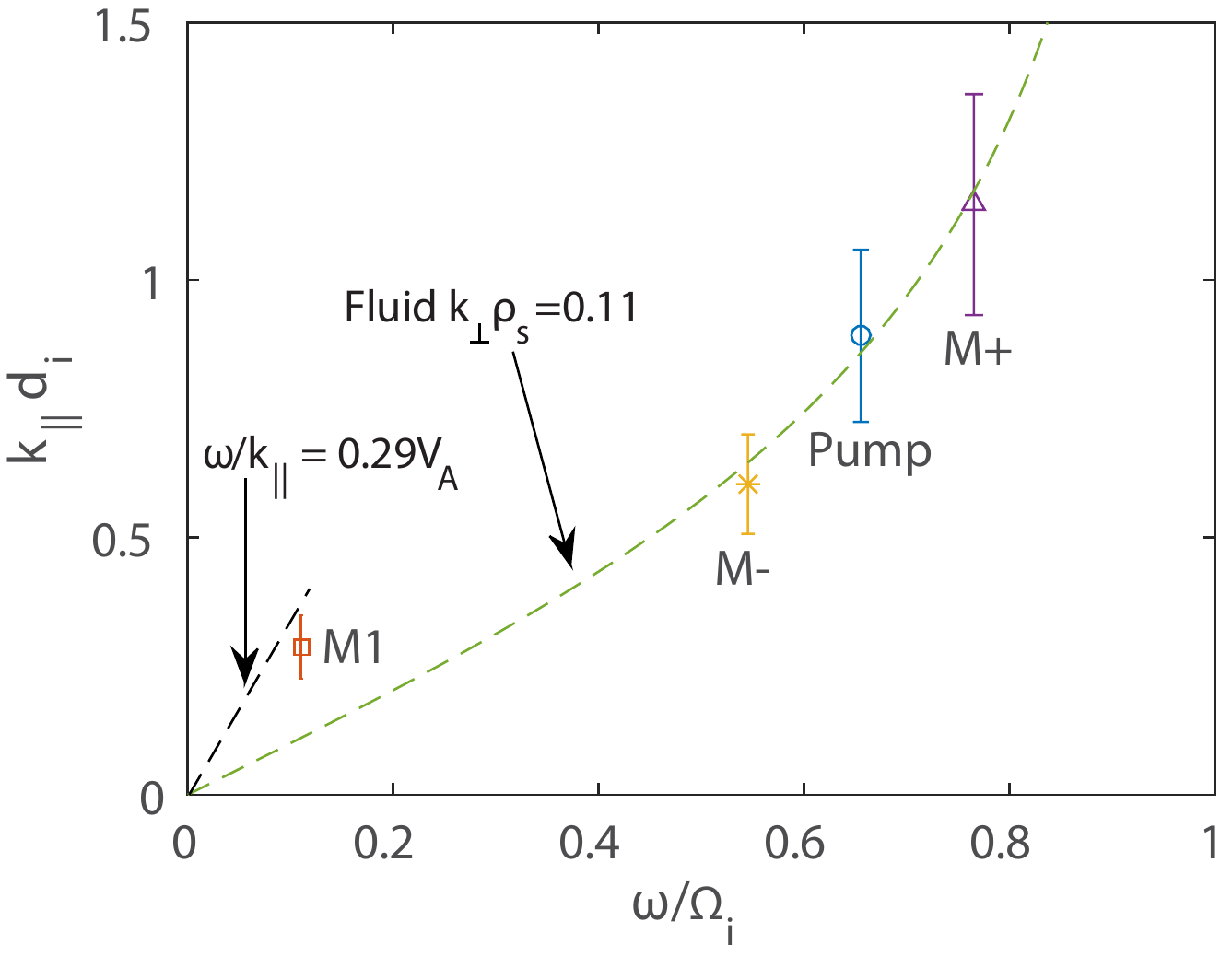}
\caption{Parallel wave number measurement showing daughter modes copropagating with the pump.  The pump, $M-$, and $M+$ are identified as KAWs while $M1$ is a nonresonant mode.  Strap antenna, $B_0=1140$~G, $\delta{B_{0\perp}}/B_0=1.9\times10^{-3}$.  Magnetic probes at $z=5.11$~m, $5.75$~m, and $6.39$~m.  The fluid dispersion relation for a KAW with the pump $k_{\perp0}\rho_s=0.11$ and a line with slope $\omega/k_{||}=0.29V_A$ are plotted for comparison.}\label{fig:decaydisp}
\end{figure}

To determine the character of the three observed daughter modes, the parallel wave numbers are measured using a set of three
axially separated magnetic probes placed $0.639$~m apart, allowing resolution of wave numbers up to $4.9/m$.  As shown in Fig.~\ref{fig:decaydisp}, this measurement reveals positive values of $k_{\parallel}$ for all modes, indicating that all three daughter modes are copropagating with the pump.  Parallel wave number matching is satisfied, $k_{\parallel\pm}\mp k_{\parallel1}=k_{\parallel0}$.  Based on the measured dispersion relation, the pump, $M-$, and $M+$ are identified as  kinetic Alfv{\'e}n waves (KAWs) while $M1$ is a nonresonant mode.  Note that $M1$ falls above the KAW dispersion curve $\omega=k_{||}V_A\sqrt{1+{\left(k_{\perp}\rho_s\right)^2}-\left({\omega / \Omega_i}\right)^2}$ for all possible values of $k_\perp$.  However, the measured $k_{||1}$ is too small for $M1$ to be an acoustic mode (for these parameters, $C_s=0.012V_A$).  This production of a nonresonant mode is consistent with the modulational instability.
    
\begin{figure}[tbp]
\centering
\includegraphics[width=\columnwidth]{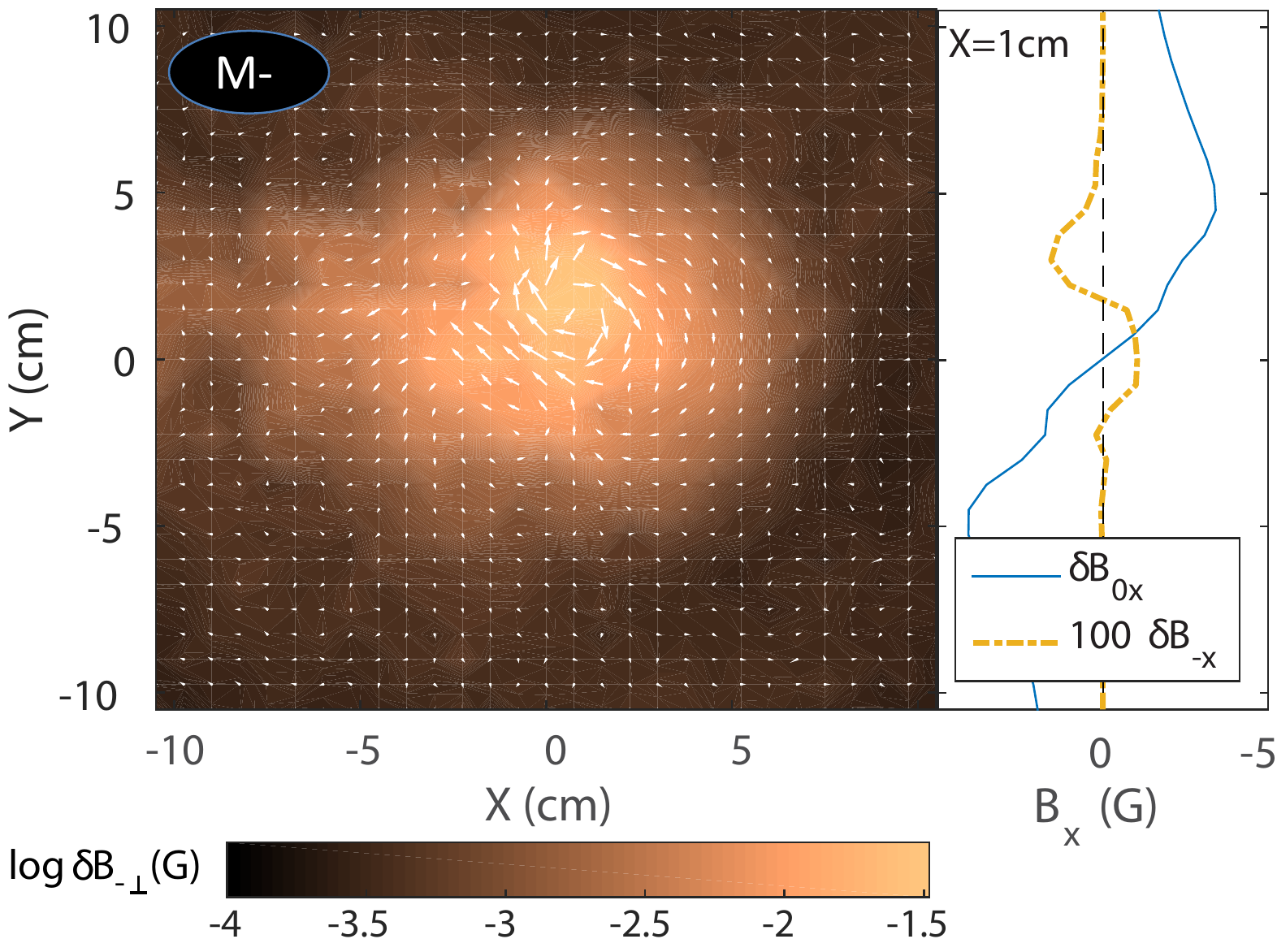}
\caption{Spatial profile of $M-$ for the strap antenna suggesting the nonlinearity is perpendicular in nature.  A cut of $\delta{B_x}$ is shown on the right.  Strap antenna pump from Fig.\ref{fig:esetup}, $B_0=1135$~G.  Color represents fluctuating magnetic field amplitude $\delta{B_{-\perp}}$; white arrows show relative magnitude and direction.  The peak in $M-$ amplitude occurs on a gradient of the pump mode magnetic field near the current channel center.}\label{fig:profiles}
\end{figure}

Measurements in the plane perpendicular to the background field reveal that perpendicular nonlinear forces likely play a role in generating the observed daughter waves.  This is shown in Fig.~\ref{fig:profiles} which displays the pattern of a representative daughter mode $M-$ in the strap antenna case; the plot is derived from a magnetic probe scanned spatially over many shots.  By comparing this figure to the strap pump mode pattern in Fig.~\ref{fig:esetup}, it can been seen that the amplitude peak of $M-$ occurs near the center of the current channel on a gradient of the pump mode magnetic field.  By contrast, the parallel ponderomotive force associated with the modulational instability will produce an amplitude peak in the daughter modes at the location where the pump wave magnetic field peaks~\citep{dorfman13a, howes13}.  This difference suggests a perpendicular nonlinearity in which perpendicular gradients of the pump mode amplitude (i.e., $k_\perp$) play a key role in the nonlinear terms.

\begin{figure}[bp]
\centering
\includegraphics[width=\columnwidth]{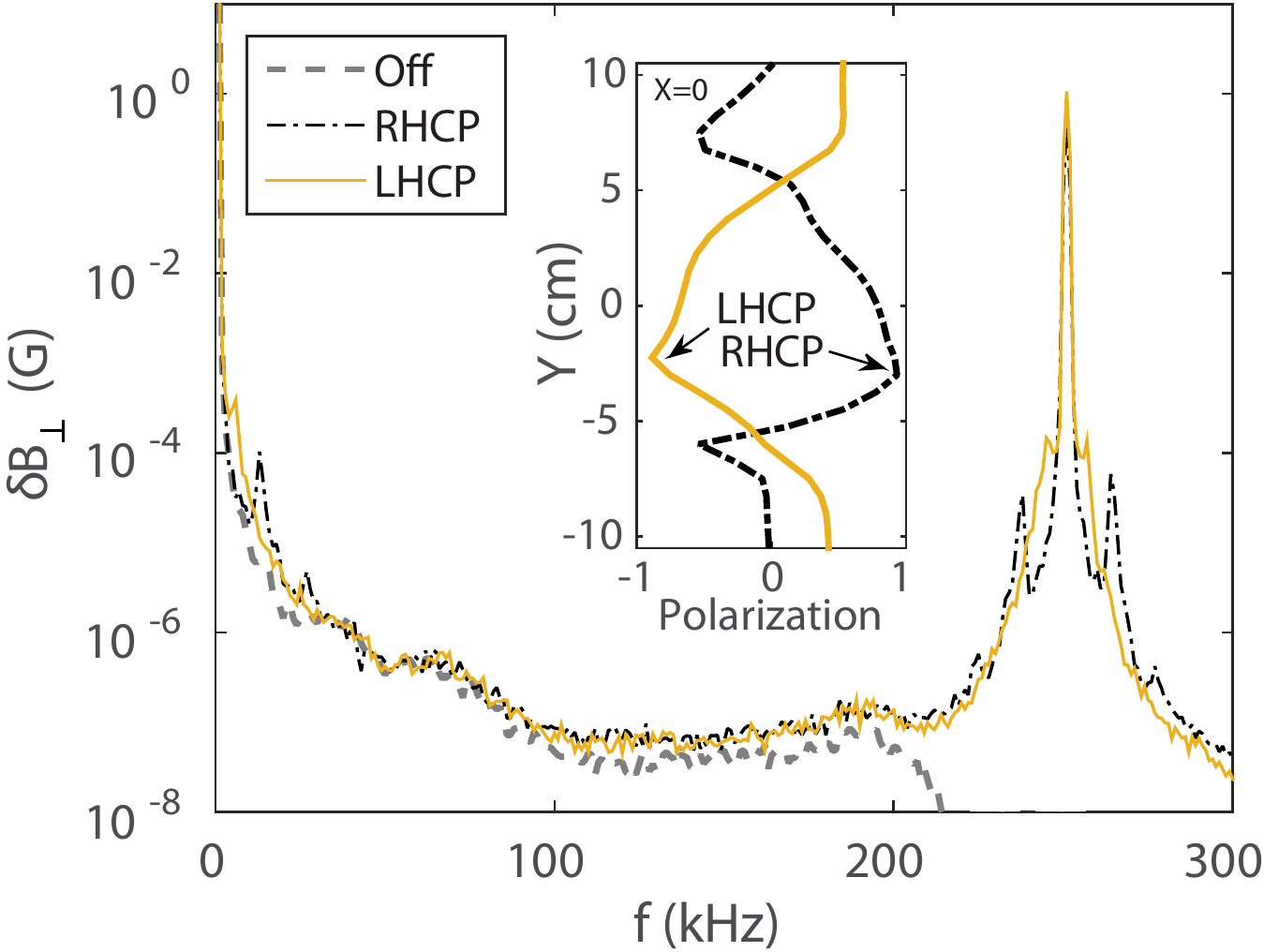}
\caption{Dependence of the observed frequency spectrum on the polarization of the RMF antenna.  Magnetic probe $x=0$, $y=-6$~cm, $z=2.6$~m.  Inset: Polarization of the RMF pump mode from Fig.~\ref{fig:esetup} along a cut at $x=0$.  $B_0=993$~G.}\label{fig:pol}
\end{figure}

\begin{figure}[tbp]
\centering
\includegraphics[width=\columnwidth]{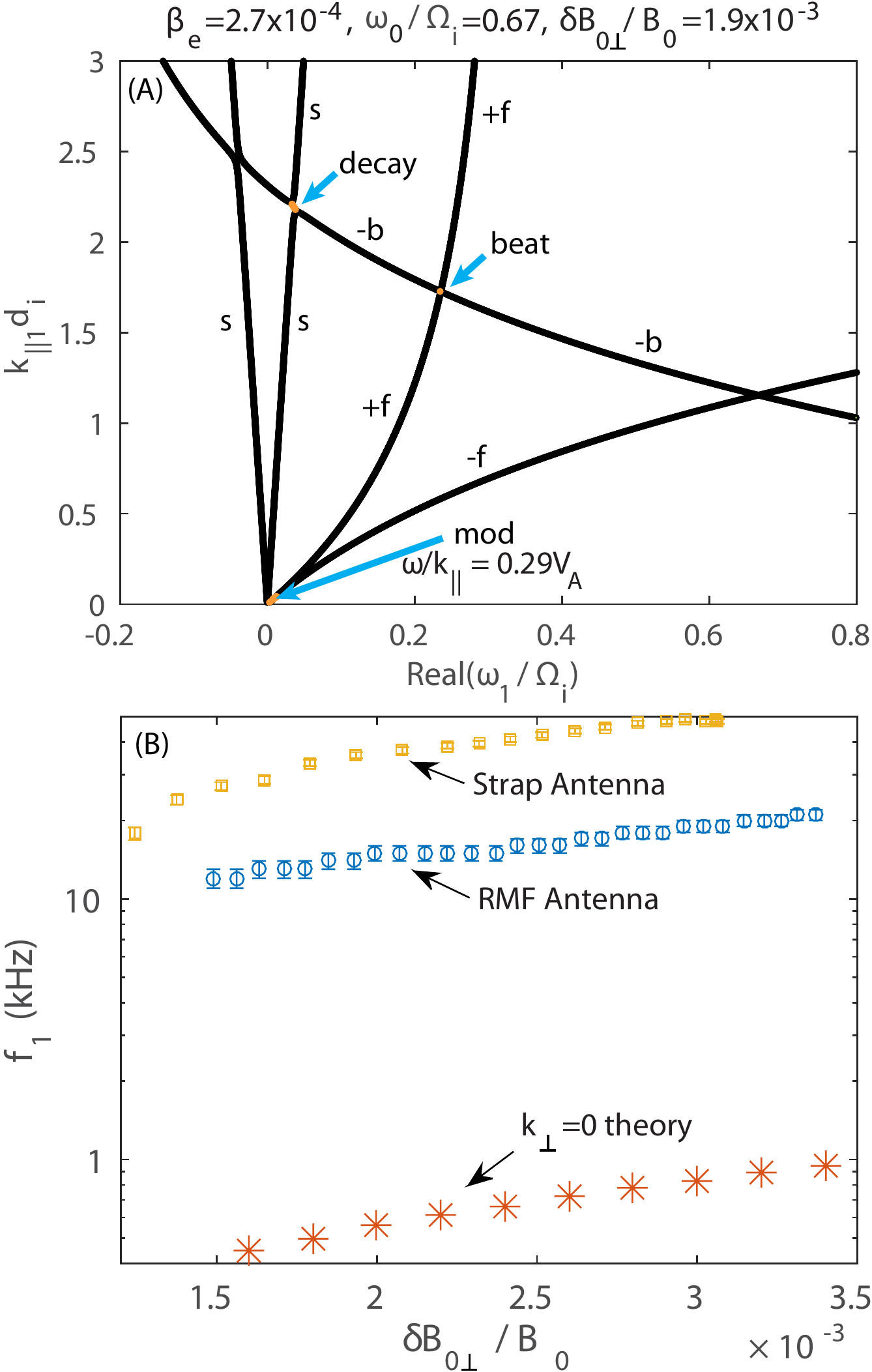}
\caption{Comparison between LAPD data and \citet{wong86} and \citet{hollweg94} $k_\perp=0$ dispersion relation.  (a) Solutions to the dispersion relation of \citet{wong86} and \citet{hollweg94} for experimental parameters of Fig.~\ref{fig:decaydisp}.  Labeled: s: sound mode, --b: backward propagating lower Alfv{\'e}nic sideband, --f: forward propagating lower Alfv{\'e}nic sideband, +f: forward propagating upper Alfv{\'e}nic sideband.  Black curves represent stable modes; orange curves representing unstable modes are labeled with the appropriate instability.  (b) Mode frequency of the modulational instability as a function of pump amplitude for experimental parameters in Fig.~\ref{fig:thpm} (blue circles), theoretical predictions (red stars), and strap antenna results with similar parameters (yellow squares).}\label{fig:disp}
\end{figure}

The pump mode polarization also influences the observed instability.  This is investigated by changing the RMF antenna phasing to produce one of the two polarization patterns shown in the inset panel of Fig.~\ref{fig:pol}.  Polarization is quantified at each spatial point by measuring the ratio of the minor to major radius in the ellipse traced by the rotating magnetic field vector.  This quantity is signed negative for left-hand rotation and positive for right-hand rotation.  As shown in Fig.~\ref{fig:pol}, left-hand (LHCP) and right-hand (RHCP) pump modes contain opposite polarization mixes that sum to linear polarization.  Each mix produces a different frequency spectra in the vicinity of the current channel; the sideband separation frequency produced by the LHCP mode is less than half that produced by the RHCP mode.  As in the linearly polarized strap antenna case, the daughter mode amplitudes peak near the current channel center for the RHCP pump mode.  The spatial profile and nonlinear physics may be different in the LHCP case and is still under investigation; the LHCP mode also leads to a broadening of the pump mode profile and a corresponding broad spectrum at low frequencies.  The existence of a polarization dependence is consistent with the theoretical literature on parametric instabilities.  However, most theoretical work (e.g., Refs.~\citep{wong86, hollweg94}) considers uniformly polarized plane waves, making direct comparisons difficult.

Despite important physical differences with the present work, modulational instability theory with $k_\perp=0$ still describes some features of the observed process well.  Figure~\ref{fig:disp}, panel~(a) shows the roots of the dispersion relation derived by \citet{wong86} and \citet{hollweg94}, solved for LAPD parameters.  This two-fluid model outputs the dispersion relation of $M1$ given a finite amplitude pump wave propagating parallel to the background field.  Orange curves for unstable modes reveal the usual decay, beat, and modulational instabilities driven by the parallel ponderomotive force.  Because the modulational instability involves only forward propogating modes, it is most consistent with the experimental observations.  An arrow on the figure indicates that the peak growth rate of the modulational instability occurs for daughter nonresonant modes with $\omega/k_{\parallel}=0.29V_A$.  Comparing this value to the measured dispersion of $M1$ in Fig.~\ref{fig:decaydisp}, the line falls just within the upper error bar.  Therefore, the fact that $M1$ is not a normal mode of the system is well predicted by modulational instability theory with $k_\perp=0$.

The theory also predicts the increase in mode frequency with pump amplitude seen in Fig.~\ref{fig:thpm}.  This is shown in panel~(b) of Fig.~\ref{fig:disp} which plots the frequency of $M1$ for both the experimental case in Fig.~\ref{fig:thpm} (blue circles) and the $k_\perp=0$ theoretical prediction~\citep{wong86, hollweg94} (red stars).  Both theory and experiment follow an upward trend.  However, the theoretical frequencies are an order of magnitude too low, and the corresponding growth times are longer than the plasma discharge; clearly, the parallel ponderomotive force is too weak to explain the experimental observations.  Furthermore, changing the $k_\perp$ spectrum of the pump wave by switching to a different antenna (yellow squares) while keeping other parameters similar results in an increase in the observed $M1$ frequency.  These observations imply that perpendicular structure plays a key role in the observed instability.

Further theoretical development is necessary to fully explain the observed daughter modes.  \citet{wong86} and \citet{hollweg94} predict that the growth rate of the decay instability should be three orders of magnitude larger than that of the modulational instability for the LAPD parameters under investigation.  Yet parametric decay to sound waves is not observed.  Possible reasons include (1) the growth rates are modified when finite $k_{\perp}$ is considered and (2) for the larger values of $k_{\parallel}$ characteristic of the decay instability ion-neutral collisions present in the experiment significantly reduce the growth rate.

Concerning the effect of finite $k_{\perp}$, very limited theoretical and computational work is available.  Numerical simulations by \citet{zanna01a}\citep{zanna15} and \citet{matteini10a} show a reduction in the growth rate of the decay instability for oblique pump waves, but do not consider the modulational instability.  Work by \citet{vinas91,vinas91a} extends the theory to allow the daughter modes to have finite $k_\perp$ while retaining $k_{\perp0}=0$ for the pump.  This allows for new classes of instabilities at oblique angles.  In particular, \citet{vinas91a} found a magnetoacoustic instability with a very narrow band of unstable wave numbers which is favored at low $\beta$ and high wave dispersion (i.e., high $\omega/\Omega_i$).  The oblique nature of the daughter modes may also explain the Alfv{\'e}nic character of the observed nonresonant mode $M1$.  New insight on the nature of the nonlinear terms may also come from extending theoretical work by \citet{brugman07} which examines copropagating waves, but only with aligned polarizations.  The applicability of these results to the present Letter is currently under investigation.

In summary, the first laboratory observations of a shear Alfv{\'e}n wave parametric instability are presented.  A single finite $\omega/\Omega_i$, finite $k_\perp$ Alfv{\'e}n wave is launched above a threshold amplitude, resulting in three daughter modes: two forward propagating Alfv{\'e}n wave sidebands and a forward propagating nonresonant mode.  Frequency and parallel wave number matching relations are satisfied.  Although these features are consistent with the $k_\perp=0$ modulational instability theory, the parallel ponderomotive force that drives this process cannot explain the growth or perpendicular spatial profile of the observed daughter modes.  Future theoretical and computational work will focus on exploring the role of $k_\perp$ in the instability.  Experimental data analysis is ongoing to explore variation with plasma parameters.

The observations reported here open a significant new avenue of research to complement extensive theory~\citep{sagdeev69,hasegawa76,goldstein78,wong86,longtin86,hollweg93,hollweg94} and simulation~\citep{ghosh93, zanna01, matteini10a, verscharen12, gao13} work on this subject.  Features of the observed instability may provide guidance to future space observation aimed at assessing the role of Alfv{\'e}n wave parametric instabilities in different regions of the heliosphere, for example, in the ion foreshock region of planetary magnetospheres where large amplitude Alfv{\'e}n waves are generated by ion beams \citep{spangler97, narita07, hoppe83}.  Because the present results are at low $\beta$, they may be of particular interest to the upcoming Solar Probe Plus mission aimed at determining what physical processes are most important in the source region of the solar wind.

\begin{acknowledgments}
The authors thank Y.~Lin, R.~Sydora, G.~Morales, and J.~Maggs for insightful discussions, S.~Vincena, P.~Pribyl, S.~K.~P.~Tripathi, and B.~Van Compernolle, for insightful discussions and assistance with the experiment, and Z.~Lucky, M.~Drandell, and T.~Ly for their excellent technical support.  S.~D.~was supported by a NASA Jack Eddy Fellowship.  This work was performed at the UCLA Basic Plasma Science Facility which is supported by DOE and NSF.
\end{acknowledgments}


\end{document}